\documentclass[fleqn,twoside]{article}
\usepackage{espcrc2}

\usepackage{graphicx}
\usepackage[figuresright]{rotating}

\newcommand{\AmS}{{\protect\the\textfont2
  A\kern-.1667em\lower.5ex\hbox{M}\kern-.125emS}}

\title{Methods of electron beam bunching}

\author{E.G.Bessonov \address[MCSD]{P.N. Lebedev Physical Institute RAS,
119991, Leninsky prospect 53, Moscow, Russia }\\}

\begin{document}

\begin{abstract}
A review of electron beam bunching methods is presented. A method to
create trains of short electron microbunches by a creation of an
electron multilayer mirror in the longitudinal phase space and then
rotating it in fields of Radiofrequency (RF) accelerators or
Free-Electron Laser (FEL) amplifiers is proposed.  \vspace{1pc}
\end{abstract}

\maketitle

               \section {Introduction}

The production of short period trains of electron bunches (electron
multilayer mirrors) is very important for the solution of the problem
of compact, stable, monochromatic, diffraction limited prebunched FELs
in the mm to X-ray regions.

Prebunched FELs were suggested by V.L.Gins\-burg in 1947 and named
"Sources of Coherent Undulator Radiation (UR)" \cite {vl}. The terms
parametric and prebunched FELs appeared later \cite {param}.  H.Motz
performed the first experiments on generation of coherent UR in 1953
\cite {motz}.  R.M.Phillips generated stimulated UR in a conventional
FEL in the RF region in 1960 \cite {phillips}. In the early 1970's
interest in prebunched FELs was renewed \cite {coh1}, \cite {coh2}.

Bunched electron beams of length $\sim 0.1 \div 1$ mm are produced
in klystron bunchers (KB) and radio-frequency accelerators \cite
{carlsten}. The basic components of a KB is an RF cavity followed by a
drift space. The cavity is excited to the $TM _{010}$ mode. The electron
beam at the exit of the buncher receives sinusoidal energy modulation.
After the drift space the electron beam will be bunched due to the
acquired velocity modulation.

A simple way to produce a bunched beam is to pass a continuous beam
through a chopper, where the beam is deflected across a narrow slit or
a system of slits, resulting in a pulsed beam behind the slit. The
chopper includes an RF cavity excited like the buncher cavity
but with the beam port offset by a distance from the cavity axis.

The chopper mode of bunching is rather wasteful. An RF buncher, which
concentrates electrons from a large range of phases towards a particular
phase is more efficient for electron bunching. However, a chopper
produces a beam with higher depth of modulation (clear bunches), does
not introduce additional energy spread and can be used for specific
reasons. Modern choppers work effectively in the RF region \cite
{wiedemann}. Schemes of choppers for harder regions can be suggested
(see below).

                    \section {Undulator bunchers}

Short-period trains of bunches are produced in undulator klystron
bunchers (UKB). The energy modulation occurs in the undulator and
electromagnetic wave fields of the UKB (FEL amplifier configuration)
and bunching takes place in a free space \cite {phillips}. The bunching
length in UKB can be shortened if a dispersion section (chicane magnet)
follows the undulator. Such a system was called an optical klystron
buncher (OKB) \cite {vinokurov}.

The wavelength of the wave used in UKB is

              \begin {equation} 
              \lambda _w = {\lambda _u \over 2 \gamma ^2_s} (1 + p
              _{\perp} ^2),
              \end {equation}
where $p _{\perp} = \sqrt {p _{\perp \,u } ^2 + p _{\perp \,w} ^2}|_{p
_{\perp \,w} \ll p _{\perp \,u}} \simeq p _{\perp \,u}$; $p _{\perp \,u
(w)} = e B _{u (w)}\lambda _{u (w)}/ 2 \pi m c ^2$ are the transverse
electron momenta determined by the electromagnetic fields of the
undulator and the electromagnetic wave; $\lambda _u$, $B _u$ are the
period and the magnetic field strength of the undulator; $B _w$, the
 magnetic field strength of the wave; $\gamma _s = 1/\sqrt {1 - \beta
_s ^2} = \sqrt {\lambda _u (1 + p _{\perp} ^2)/4 \lambda _w}$, the
relative electron equilibrium energy; $\beta _s = \sqrt {\beta _{\perp
\,s}^2 + \beta _{z \,s} ^2}$; $\beta _{\perp \,s} = \beta _{\perp\,u
\,s} + \beta _{\perp\,w \,s} = (p _{\perp \,u} + p _{\perp \,w})/\gamma
_s $; $\beta _{z \,s} = \lambda _u/ (\lambda _u + \lambda _w)$; $2\pi
mc^2 /e \simeq 10700$ $ Gs \cdot cm$; $e, m$, the electron charge and
mass \cite {alf-bess}.

Undulators with high deflection parameters (wigglers) possess high
dispersion. If the UKB works in the regime of high deflection parameter
and high intensity of the wave (HHKB), the energy modulation and
bunching can occur in the same undulator \cite {alf-bess}. In HHKB a
high degree of bunching occurs if electrons produce a quarter of a
period of phase oscillations. The bunching length in this case can be
short ($\sim 1 $ m in the optical region, see section 3).

A scheme of bunching based on convergent waves and/or undulators with
variable parameters (VPKB) was developed in \cite {bes-ser}. In this 
scheme an electron beam is trapped as a whole in regions less than 
those limited by separatrices of the FEL amplifier for about 3/4 of a 
period of phase oscillations (see Fig. 1). Clear bunched beam can be 
used for high-efficiency generation on the fundamental or higher 
harmonics in the same undulator or in an undulator radiator. A similar 
scheme of bunching can occur in cases being developed in \cite 
{neil}-\cite {{neil-merm}}.

\vskip 25mm
\hskip 1mm
\begin{picture}(50,40)
\includegraphics[scale=0.5]{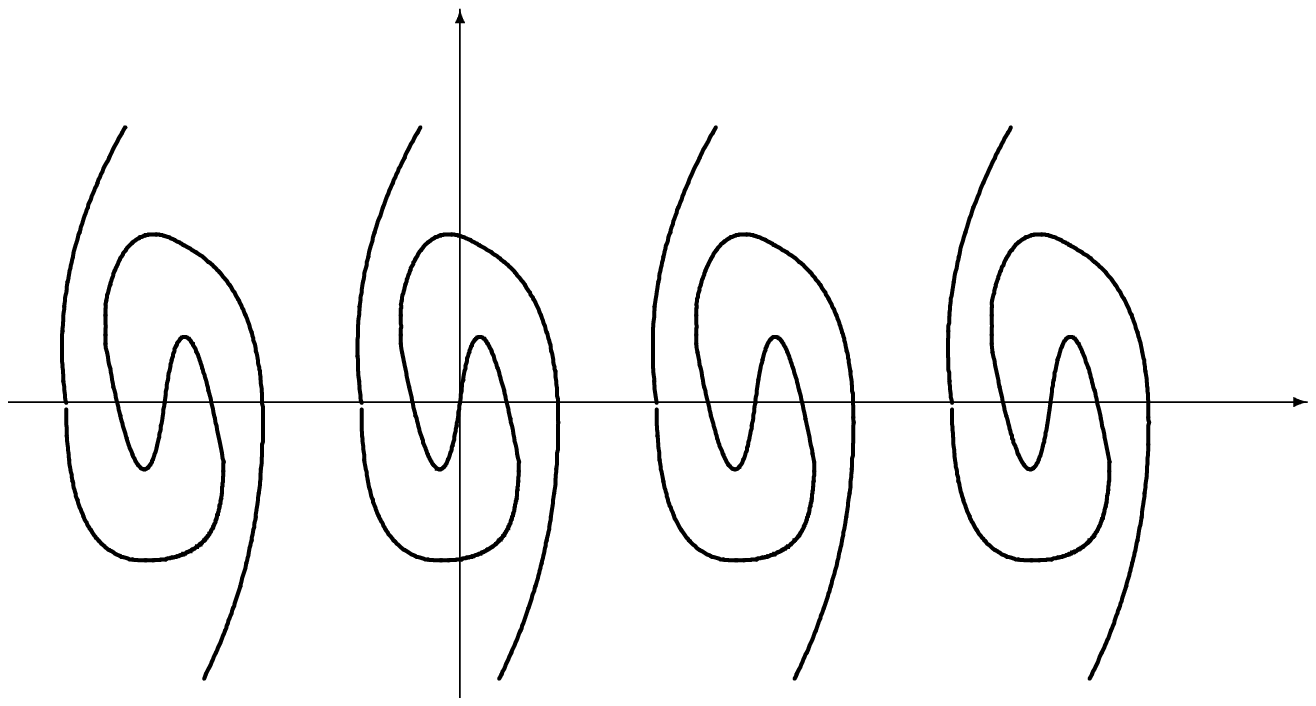}
\put(-12, 37){\small x}
\put(-120, 100){\small $\varepsilon - \varepsilon _0$}
\end{picture}
\vskip -0mm 
\noindent {\small Fig. 1: Electron beam at the exit of VPKB in the
phase space ($\varepsilon - \varepsilon _0$, $x$). The amplitude of the
laser electric field is $E _{w \, m} = E _0 (1 + \alpha z)$. $E _0 =
180$ kV/cm, $\alpha = 0.8$ m$^{-1}$, $\lambda _u = 1.5$ cm, the
undulator length $L = 5$ m, $B _u = 3.10 ^3 Gs$, $\lambda _w = 1.06$
mkm.}

\vskip 5mm

A bunching scheme based on extraction of a series of microbunches
from an electron beam located in storage ring buckets when an undulator
with variable parameters and electromagnetic wave are used was
considered in \cite {bess99}. Extracted microbunches stay in the
storage ring buckets.

Shock acceleration of electron bunches or trains of bunches by
wavepackets with a longitudinal component of electric field can lead to
their compression. Such schemes can be realized in linear accelerators
\cite {bks76} and laser wake-field accelerators.

Below, a method of electron-beam bunching and schemes of
realization are considered.

        \section {A Method of Electron Beam Bunching}

The proposed method is based on production of electron beams in the
form of trains of electron bunches, composed in the longitudinal plane
in $N$ energy layers and followed by rotation of the bunches at a right
angle \cite {bess95}. Rotation converts the bunches into short-period
($cT _{mb} < \lambda _L /N$) trains of micro bunches with layers
transverse to the beam propagation (see Fig. 2).

\vskip 22mm
\hskip -5mm
\begin{picture}(50,40)
\includegraphics[scale=0.5]{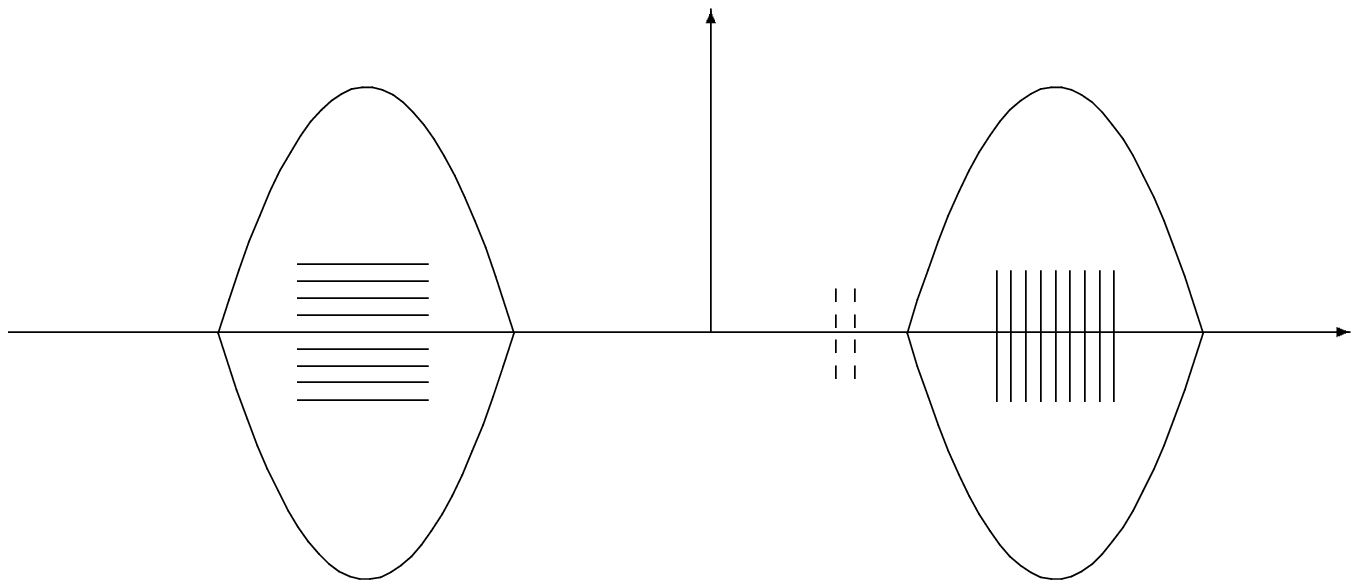}
\put(-12, 37){\small x}
\put(-92, 90){\small $\varepsilon$}
\end{picture}
\vskip -0mm 
\noindent {\small Fig. 2: An electron bunch arranged in energy
layers in the left bucket is converted after its rotation in the right
bucket to a train of micro bunches.}

\vskip 5mm

The longitudinal dimension of microbunches will be determined by the
energy spread of layers and non linearity of phase oscillations. The
influence of the non linearity can be neglected if the phase length of
layers is short ($ < \pi$).

      \subsection {Arrangement of electron beams with energy layers}

{\it Example 1}. A system of electron beams produced by $N$ electron
guns of energy $\varepsilon _n = \varepsilon _0 + n \Delta
\varepsilon _l$ can be arranged in one beam with energy layers by means
of a bending magnet (see Fig. 3) \cite {bess95}.  Here $n = 0, 1, 2,
3,...  N - 1$, $\Delta \varepsilon _l$ is the interval between energy
layers.

\vskip 3mm
{\it Example 2}. A thin electron beam with finite energy spread and low
transverse emittance passes through a dispersive magnet and a metal
lattice with $N$ slits (see Fig. 4).  After the lattice, the electron
beam is arranged in $N$ energy layers\footnote {More perfect systems of
focusing bending magnets and quadrupole lenses can be used in this
scheme.}.

If the electron beam has a small energy spread and an energy
modulation, then this scheme can work in a chopper mode at RF, optical
and harder regions.

If the ionization losses of electrons in the medium of a thin lattice
are equal to $\Delta \varepsilon _l/2$, then electrons of the incoming
beam will be arranged in layers without losses.

If electron RF guns in Examples 1,2 produce short bunches, the beam at
the exit of systems will be arranged in layers and bunched. The beam
can be bunched at the exit of the bending magnet as well. After
rotation at the bunching frequency, the beam will be a train of bunches
a buncher's wavelength apart. Bunches will consist of $N$ microbunches
(see Fig. 2).

\vskip 30mm
\hskip 5mm
\begin{picture}(50,40)
\includegraphics[scale=0.65]{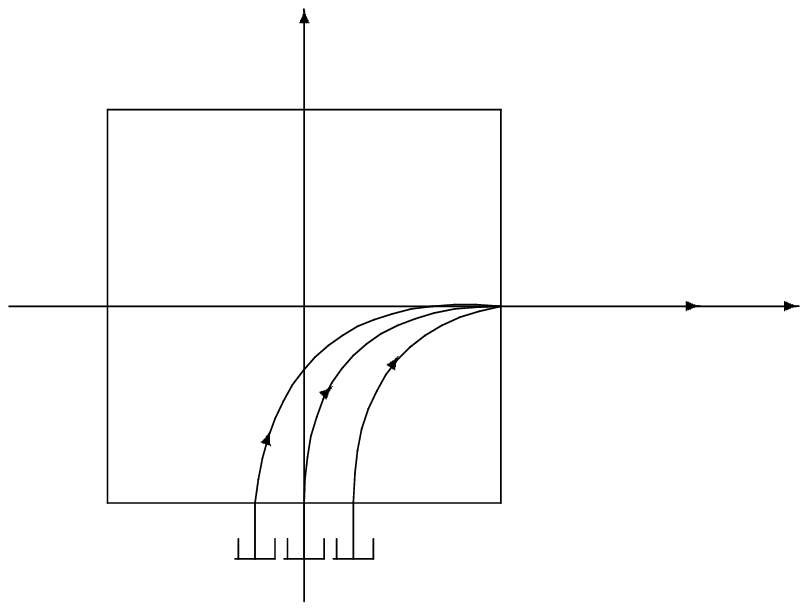}
\put(-20, 57){\small x}
\put(-100, 115){\small $z$}
\put(-98, 0){\small $G _3$}
\put(-109, 0){\small $G _2$}
\put(-120, 0){\small $G _1$}
\put(-90, 90){\small $BM$}
\put(-90, 45){\small e}
\put(-103, 40){\small e}
\put(-112, 33){\small e}
\put(-40, 57){\small e}
\end{picture}

\vskip 1mm 
\noindent {\small Fig. 3: The scheme of arranging an electron beam in
energy layers. $G _1$, $G _2$, $G _3$ are the electron guns, $BM$ is
the bending magnet.}

\vskip 32mm
\hskip 5mm
\begin{picture}(50,40)
\includegraphics[scale=0.65]{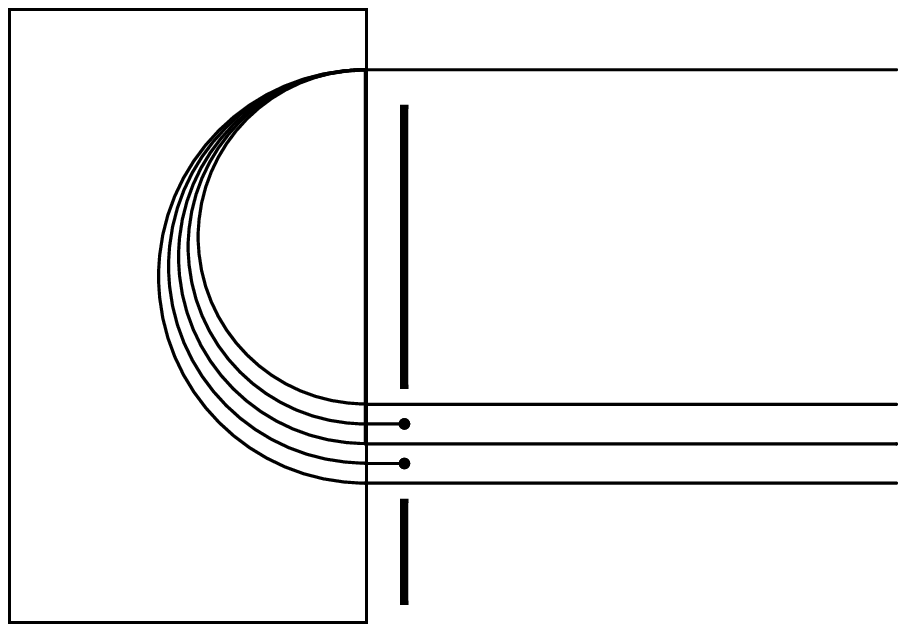}
\put(-210, 90){\small $BM$}
\put(-94, 47){\small $\varepsilon _1$}
\put(-103, 38){\small $\varepsilon _2$}
\put(-112, 29){\small $\varepsilon _3$}
\put(-90, 110){\small $e-beam$}
\put(-140, 55){\small Lattice}
\put(-76, 41){\small $>$}
\put(-85, 33){\small $>$}
\put(-94, 26){\small $>$}
\put(-110, 104){\small $<$}
\end{picture}

\vskip 2mm 
\noindent {\small Fig. 4: The scheme of production of the electron
beam arranged in layers. $BM$ is the bending magnet.}

\vskip 3mm

{\it Example 3}. Two counter-propagating short laser beams stored in an
open resonator perform an interference lattice in a storage ring at a
position of low-beta and high dispersion function. A low-emittance
electron beam intersects the lattice at a 90 degree angle \cite
{shintake}.

In this case, the energy losses of electrons depend on their position in
the laser beam. Electrons moving in the bright zones of fringes lose
more energy than those in the dark zones. Their energy (closed orbits)
will tend to the energy (closed orbits) of electrons in the nearest
dark zones if the energy of scattered quanta $\hbar \omega _{max} \ll
\Delta \varepsilon _l$. As a result, the electron beam will be arranged
in energy layers if the interaction time will be much less than the
period of phase oscillations of electrons in the storage ring.

More effective interaction of counterpropagating laser and electron
beams can be produced in straight sections of storage rings if
$TEM _{n, \, 0}$ laser mode is used ($n > 1$).

This scheme of bunching can be realized in ion storage rings using
present-day technology (Rayleigh cross-section is $7 \div 10$ orders
higher than Compton one) \cite {bess95}. Realization of the electron
version of this scheme depends on the possibility of storing high-power
laser beams in high-finesse optical resonators\footnote {Note that
prebunched Free-Ion Lasers present the ultimate in the capabilities
of lasers \cite {nim95}. Relativistic ion beams can be cooled to a
high degree in 6D space \cite {prl}.}.

     \subsection {Rotation of electron beams in an FEL amplifier}

The frequency of phase oscillations of electrons in a helical undulator 
and a circular polarized laser wave $\omega _{\varphi} = \omega 
_{u}\sqrt {B _u B _w (1 + \beta _{z\,s})}/\beta _{z\, s} \gamma _s B 
_c$, where $\omega _u = 2\pi \beta _{z \,s} c/\lambda _u$; $B _c = 2 
\pi m c^2 /e \lambda _u$ \cite {alf-bess}\footnote {We suppose that the 
laser beam intensity is constant along the interaction region of the 
laser and electron beams ($l _{1/4} < 2 l _R$, where $l _R = 4\pi 
\sigma _L ^2 /\lambda _L$ is the Rayleigh length).}.

Electrons produce a quarter of the period of phase oscillations inside
the length

              $$l _{1/4} = {\pi c \beta _{z\, s} \over 2 \omega
              _{\varphi}} = {\beta _{z \,s} \gamma _s \over 4} \sqrt
              {\lambda _u \lambda _w \over (1 + \beta _{z \,s}) p
              _{\perp \,u} p _{\perp \, w}} $$

                     \begin {equation} 
              = {\lambda _{u} \over 8 \sqrt 2}\sqrt {1 + p _{\perp}
              ^2 \over p _{\perp \,u} p _{\perp \, w}}.
                     \end {equation}

The maximum deviation of the electron energy in the bucket determined
by a separatrix is

             $$ {\Delta \gamma _{sep} \over \gamma _s} = {2 \beta _
              {z\, s} \omega _{\varphi}\over (1 + \beta _{z\, s})
              \omega _u} $$
                     \begin {equation} 
              = 4 \beta {z \,s} \sqrt {p _{\perp \,u} p
              _{\perp \, w} \over (1 + \beta _{z \,s}) (1 + p
              _{\perp} ^2)}.
                     \end {equation}

The magnetic field strength of the circular polarized Gaussian laser
beam

                     \begin {equation} 
          B _{w}[Gs] = \sqrt {2 P _{w} \over c \sigma _{l} ^2} \simeq
          2.58 \cdot 10 ^{-2} \sqrt {P[W] \over \sigma _l ^2 [cm ^2]}.
                     \end {equation}

{\it Example 4.} The parameters of the undulator and electromagnetic
laser wave are the following:

\vskip 3mm
\noindent laser power \hfill $P _L = 1.6 \cdot 10 ^{9}$ $W$,

\noindent laser wavelength \hfill  $\lambda = 10 ^{-4}$ $cm$,

\noindent laser beam dispersion \hfill $\sigma _{L} = 2\cdot 10 ^{-2}$
$cm$,

\noindent undulator  magnetic field \hfill $B _u = 10700$ $Gs$,

\noindent undulator period  \hfill $\lambda _u = 3$ $cm$,

\noindent deflecting parameter \hfill  $p _{\perp} = 3$,

\noindent Duration of laser wavepacket \hfill $\tau _l = 2.5 \cdot 10
^{-11} sec$,

\noindent energy of laser wavepacket \hfill $\varepsilon _l =
P_l \tau _l = 0.04$ J,

\vskip 4mm
In this case $\gamma _s = 274$, $p _{\perp \, u} = 3$, $p
_{\perp \, w} = 4.82 \cdot 10 ^{-3}$, $l _{1/4} = 69.7$ $cm$, $B
_w = 5.16 \cdot 10 ^5$ $Gs$, $l _R = 1.0$ $m$, ($2 l _R > l
_{1/4}$), $\Delta \gamma _{sep}/\gamma _{s} = 1.07 \cdot 10
^{-2}$.

                    \section {Conclusion}

Linear accelerator, storage ring and buncher technologies offer
high-beam quality and short bunches necessary for operation of high
efficiency, high-power, high-degree monochromatic prebunched FELs in
optical and X-ray regions.

This work was supported partly by the Russian Foundation for Basic
Research, Grant No 02-02-16209.

\end{document}